\documentclass[graybox]{svmult}

\usepackage{mathptmx}       
\usepackage{helvet}         
\usepackage{courier}        
\usepackage{type1cm}        
\usepackage{makeidx}         
\usepackage{graphicx}        
\usepackage{multicol}        
\usepackage[bottom]{footmisc}
\usepackage{natbib}

\makeindex             

\newcommand{\Msun}{\rm{M}_{\odot}}

\begin{document}

\title*{Luminosity functions in the CLASH-VLT cluster MACS\,J1206.2-0847: the importance of tidal interactions}
\titlerunning{LFs in M1206}
\author{A. Mercurio$^{1}$, M. Annunziatella$^{2,3}$, A. Biviano$^{3}$, M. Nonino$^{3}$, P. Rosati$^{4}$, I. Balestra$^{3}$, M. Brescia$^{1}$, M. Girardi$^{2,3}$, R. Gobat$^{5}$, C. Grillo$^{6}$, M. Lombardi$^{7}$, B. Sartoris$^{2,3}$, and the CLASH-VLT team}
\authorrunning{Mercurio et al.}
\institute{\email{mercurio@na.astro.it}.
\footnotesize{\at $^1$ INAF/Oss. Astronomico di Capodimonte, via Moiariello 16, 80131, Napoli, Italy, \at $^2$ Dip. di Fisica, Univ. di Trieste, via Tiepolo 11, 34143 Trieste, Italy, \at $^3$ INAF/Oss. Astronomico di Trieste, via G. B. Tiepolo 11, 34143, Trieste, Italy, \at $^4$ Dip. di Fisica e Scienze della Terra, Univ. di Ferrara, via Saragat 1, 44122, Ferrara, Italy, \at $^5$ Korea Institute for Advanced Study, KIAS, 85 Hoegiro, Dongdaemun-gu Seoul 130-722, Republic of Korea, \at $^6$ Dark Cosmology Centre, Niels Bohr Inst., Univ. of Copenhagen, Juliane Maries Vej 30, 2100 Copenhagen, Denmark, \at $^7$ Dipartimento di Fisica, Università degli Studi di Milano, via Celoria 16, I-20133 Milan, Italy.}}
%
%
\maketitle

\abstract*{}

\abstract{We present the optical luminosity functions (LFs) of
  galaxies for the CLASH-VLT cluster MACS\,J1206.2-0847 at z=0.439,
  based on HST and SUBARU data, including $\sim$ 600 spectroscopically
  confirmed member galaxies. The LFs on the wide SUBARU FoV are well
  described by a single Schechter function down to M$\sim$M$^*$+3,
  whereas this fit is poor for HST data, due to a faint-end upturn
  visible down M$\sim$M$^*$+7, suggesting a bimodal behaviour. We also
  investigate the effect of local environment by deriving
  the LFs in four different regions, according to the distance from
  the centre, finding an increase in the faint-end slope going from
  the core to the outer rings. Our results confirm and extend 
    our previous findings on the analysis of mass functions,
  which showed that the galaxies with stellar mass below $10^{10.5}\,
  \Msun$ have been significantly affected by tidal interaction
  effects, thus contributing to the intra cluster light (ICL).}

\section{Introduction}
\label{sec:1}

The galaxy LF, which describes the number of galaxies per unit volume
as function of luminosity, is a powerful tool to study the properties
of galaxy populations and to constrain their evolution through
comparisons with the dark matter halo mass function. The
dependence of the observed galaxy LF on the environment provides a
powerful discriminator among the environmental-related mechanisms that have been suggested to drive galaxy transformations,
e.g. merging, ram-pressure stripping, tidal
interactions. Interestingly, in dense environments, a single Schechter
function was found to be a poor fit of the LF, due to the presence of
an upturn at fainter magnitudes (e.g. \citealt{agu14} and references
therein). This feature is present in dynamically evolved regions
(i.e. regions with a large fraction of elliptical galaxies, a high
galaxy density, and a short crossing time) but absent in unevolved
regions, such as the Ursa Major cluster and the Local Group
(e.g. \citealt{tre02}). The LFs can vary from cluster to cluster
or/and also in different cluster regions, depending on the mixture of
different galaxy types induced by cluster-related processes. In order
to asses the relative importance of the processes that may be
responsible for the galaxy transformations, we have performed a
photometric study of the cluster MACS\,J1206.2-0847 at z=0.439, by
examining the effect of local environment through the
comparison of the LFs in four cluster regions.  This study has an
unprecendeted combination of depth and area, taking advantage of both
multiband HST data and Subaru panoramic imaging, as well as extensive
spectroscopic coverage.

\section{Data and catalogues}
\label{sec:2}

Ground-based photometric observations were carried out with the
SuprimeCam at Subaru, covering 30$^\prime$ $\times$ 30$^\prime$, with
total exposure times of 2400\,s, 2160\,s, 2880\,s, 3600\,s and 1620\,s
in B, V, R$_{\mathrm{c}}$, I$_{\mathrm{c}}$ and z band respectively,
and seeing values between 0.7 and 1.0\,arcsec.
The photometric catalogues were extracted using the software
SExtractor (\citealp{ber96}) in conjunction with PSFEx
(\citealp{ber11}), which performs PSF fitting photometry.
Spectroscopic data were acquired with the VIMOS instrument at the ESO
VLT, as part of the ESO Large Programme CLASH-VLT {\it Dark Matter
Mass Distributions of Hubble Treasury Clusters and Foundations of
$\Lambda$CDM Structure Formation Models} (P.I. Piero Rosati; see
\citealp{ros14}).

The final photometric catalogue contains $\sim$ 34000 objects down to
$\mathrm{R_c} \, =\, 24$ mag. The CLASH-VLT campaign provided 2749
reliable spectroscopic redshifts. To complete the spectroscopic
sample, we have used photometric redshifts, calibrated on the large
redshift catalog, as described in Biviano et
al. \citeyear{biv13}. Thus, our final sample includes 2468 cluster
members (590 of which are spectroscopically confirmed). In order to
derive the LFs, we also applied the completeness corrections to the
observed galaxy counts reported in \citealt{ann14}.

MACS\,J1206.2-0847 was observed with HST in 16 broadband filters, from
the UV to the near-IR as part of the CLASH multi-cycle treasury
program (see Postman et al.~\citeyear{pos12}). As described in Grillo
et al. \citeyear{gri15}, in the HST/WFC3 FoV we selected cluster
members by measuring the probability for a galaxy to be cluster member
according to its multi-dimensional color space distribution from 12
CLASH bands (excluding the F225W, F275W, F336W, and F390W bands, due
to the low signal-to-noise), based on the color distribution of  the 233 galaxies in the spectroscopic sample. With this method, we
obtained a total sample of 1177 spectroscopic or photometric members, down
to F814W=26.5 mag.

This dataset has allowed us to investigate the cluster galaxy
population down to M$\sim$M$^*$+7 (corresponding to stellar masses of
$\mathcal{M}\sim$ 10$^{8.2}$ $\Msun$) in the central region
(R$<$0.4\,r$_{200}$) and M$\sim$M$^*$+3 ($\mathcal{M}\sim$10$^{9.5}$
$\Msun$) out to $3.5\,r_{200}$ (r$_{200}$=1.96 Mpc; \citealp{biv13}).

In order to match the photometry of the HST data in the F814W band, and of the SUBARU data in the Rc band, we derived a linear relation between the magnitudes measured in these two bands

In order to match the photometry of the HST data in the F814W
band and of the SUBARU data in the $\mathrm{R_c}$ band, we derived a linear relation between the magnitudes measured in these two bands on.

\section{Results}
\label{sec:3}

We show in Fig.~1 (left panel) the R$_{\mathrm{c}}$ LFs obtained from
SUBARU data in four different cluster environments. We differentiate
the environment in the cluster with the clustercentric distance from
the brightest cluster galaxy. As reported in \citealt{ann14}, this is
equivalent to referring to the local galaxy number density in this
dynamically relaxed cluster (but see also \citealt{gir15}). The LFs
are well described by a single Schechter function. Environmental
effects are apparent in the slope of LFs, which is found to be steeper
in the outskirts compared to the central region, by more than
1$\sigma$. We assess the statistical significance of the
  difference between LFs non parametrically, via a Kolmogorov-Smirnov
  test, which gives a very low probability ($<$0.04\% in the better
  case) to the hypothesis that the distribution of galaxies are drawn
  from the same population in the first region respect to the three
  adiacent areas. This steepening is also in agreement with the
stellar mass function (MF) presented by \citealt{ann14}.  As shown in
Fig.~1 (right panel), we find a deficit of low mass galaxies in the
cluster core, which adds up to $\sim$ 6$\times$10$^{11}$ $\Msun$,
computed by extrapolating the slope of the MF in the external
region. Interestingly, this stellar mass matches the ICL stellar mass
estimated by \citealt{pre14}. Thus, the deficit of these galaxies in
the core can be interpreted as stripped stellar mass which went to
populate the ICL component during the cluster assembly process.
The deficit of faint galaxies could be also related to galaxy
  merging processes, however, we don't observe an excess of massive
  galaxies in cluster cores compared to more external regions (see
  \citealt{ann14}). In Fig.~2, we present the LF of the cluster
members in HST data (in black), compared with that obtained from
SUBARU data in the same region (in red) and in the outer region (in
blue). Ground-based photometric data are adequately fitted by a single
Schechter function.  Instead, as for the central $\sim$0.4 r/r$_{200}$
region mapped by HST, the fit with a single Schechter function is
poor. The residuals show that the fit systematically under- and
over-predicts the observed counts in the range
21$<$R$_{\mathrm{c}}<$23.5 ($10^{10.5}<\mathcal{M}< 10^{9.5}\,\Msun$),
and there is an upturn at R$_{\mathrm{c}}>$23.5, suggesting that a
bimodal behaviour of the LF is more suitable to describe the LF down
to M=M$^∗$+7. This result confirms the scenario reported above for the
ICL and supports the idea that the ICL is built-up by the tidal
stripping of $\sim$M$^*$+2 mag galaxies (\citealt{dem15}). We are now
extending the analysis of the LF and MF to other clusters from the
CLASH-VLT dataset, three of which include $\sim\! 1000$ spectroscopic
members. This high-number statistics will allow us to also take into
account structural and stellar population properties of the cluster
members, to elucidate the impact of tidal effects in the mass assembly
history of massive clusters.

%
\begin{figure*}[b]
\vspace{-0.6cm}
\begin{minipage}{0.5\textwidth}
\centering
\hspace{-0.4cm}
\includegraphics[scale=.23,angle=-90]{fig1a.ps}
\label{fig1a}
\end{minipage}\hfill
\begin{minipage}{0.5\textwidth}
\centering
\hspace{-1.0cm}
\includegraphics[scale=0.3]{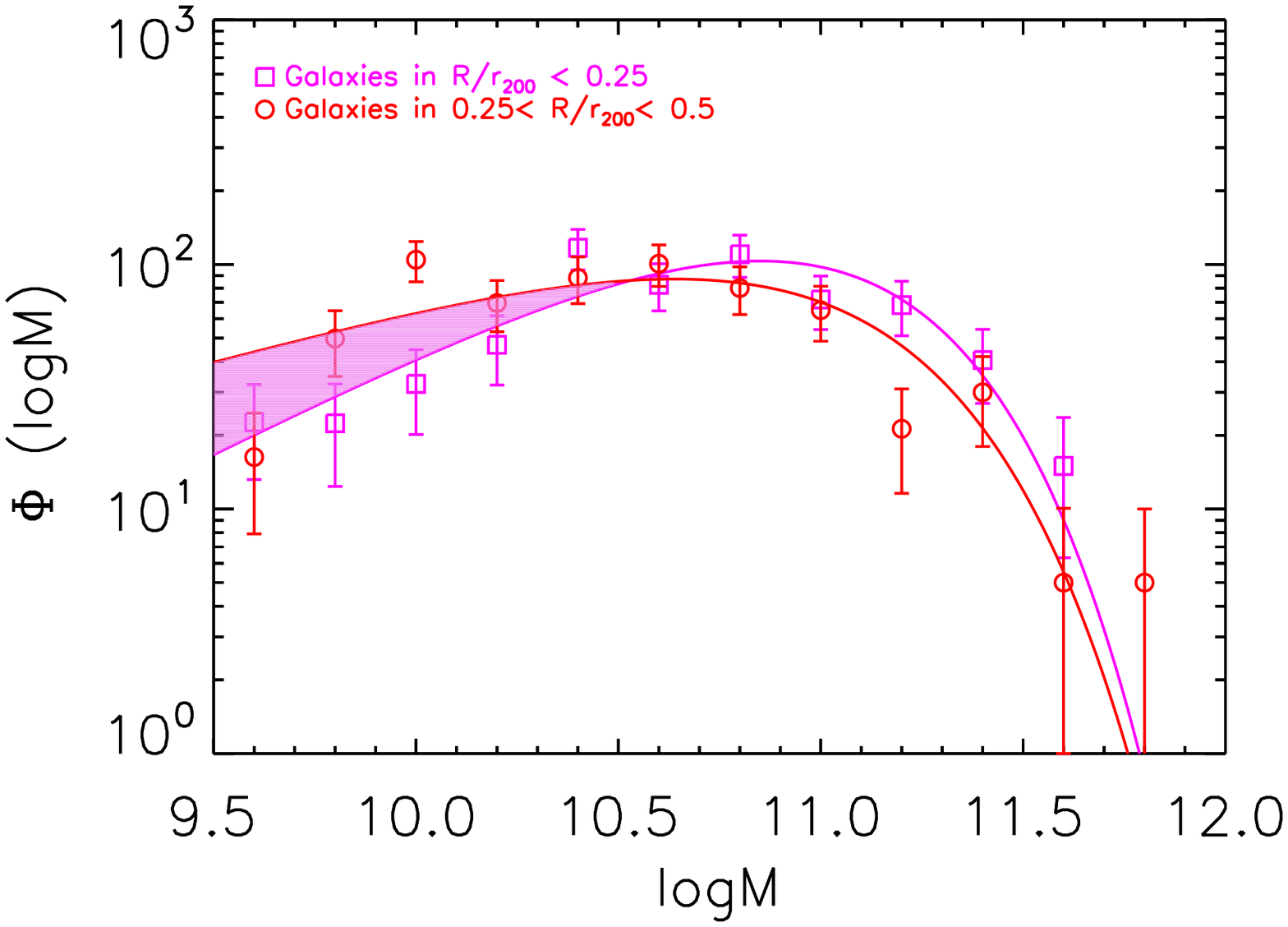}
\label{fig1b}
\end{minipage}
\caption{{\it Left panel}: LFs of galaxies in MACS\,J1206.2-0847,
  covering regions with increasing distance from the
  centre. Continuous lines are the fits with the Schechter
  function. The best--fit Schechter parameters for $\alpha$ and M$^*$
  with their 1$\sigma$ contours for the corresponding LFs are also
  reported in the small panel. {\it Right panel}: MFs of galaxies in
  the core of MACS\,J1206.2-0847 and the outer adjacent ring. The
  shaded area highlights the deficit of low mass galaxies in the core,
  which adds up to $\sim 6 \times 10^{11} \Msun$ (see \citealt{ann14}
  for details).}

\end{figure*}

\begin{figure}
\begin{minipage}{0.5\textwidth}
\centering
\caption{LFs of galaxies in MACS\,J1206.2-0847, from HST data (black)
  and from SUBARU data in the same region covered by HST (red) and in
  the external region (blue). Continuous lines are the fits with the
  Schechter function. Dashed line is the fit to HST data down to
  M=M*+3. The best--fit Schechter parameters for $\alpha$ and M$^*$
  with their 1$\sigma$ contours for the corresponding LFs are also
  reported in the small panel.}
\end{minipage}\hfill
\begin{minipage}{0.5\textwidth}
\centering
\includegraphics[scale=.23,angle=-90]{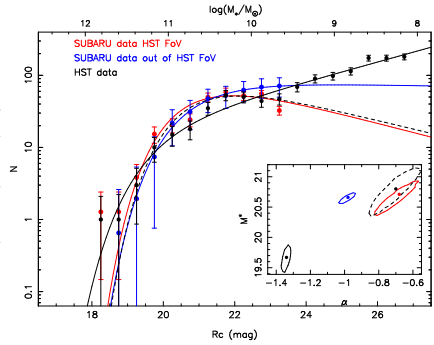}
\end{minipage}
\label{fig:2}       
\end{figure}

\begin{acknowledgement}
The authors acknowledge financial support from PRIN-INAF 2014: {\it Glittering
Kaleidoscopes in the sky, the multifaceted nature and role of galaxy
clusters} (PI M. Nonino).
\end{acknowledgement}

{}

\begin{thebibliography}{}
{\footnotesize
\bibitem[\protect\citeauthoryear{Agulli et al.}{2014}]{agu14}
Agulli I., Aguerri J. A. L., S\`{a}nchez-Janssen R., et al. 2014, MNRAS, 444, 34

\bibitem[\protect\citeauthoryear{Annunziatella et al.}{2014}]{ann14}
Annunziatella M., Biviano A., Mercurio A., et al. 2014, A\&A, 571, A80

\bibitem[\protect\citeauthoryear{Bert\`{\i}n \& Arnouts}{1996}]{ber96}
Bert\`{\i}n E. \& Arnouts S. 1996, A\&AS, 117, 393

\bibitem[\protect\citeauthoryear{Bert\`{\i}n}{2011}]{ber11}
Bert\`{\i}n E. 2011, ASP Conference Series, 442, 393

\bibitem[\protect\citeauthoryear{Biviano et al.}{2013}]{biv13}
Biviano A., Rosati P., Balestra I., et al. 2013, A\&A, 558,1

\bibitem[\protect\citeauthoryear{DeMaio et al.}{2015}]{dem15}
DeMaio T., Gonzalez A., Zabludoff A., Zaritsky D., Bradac M., 2015, MNRAS,  arXiv:1501.02225

\bibitem[\protect\citeauthoryear{Girardi et al.}{2015}]{gir15}
Girardi M., Mercurio A., Balestra I., et al. 2015, A\&A, arXiv:1503.05607

\bibitem[\protect\citeauthoryear{Grillo et al.}{2015}]{gri15}
Grillo C., Suyu S. H., Rosati P., et al. 2015, ApJ, 800, 38

\bibitem[\protect\citeauthoryear{Postmann et al.}{2012}]{pos12}
Postmann M., Coe D., Benitez N., Bradley L., et al. 2012, ApJS, 199, 25

\bibitem[\protect\citeauthoryear{Presotto et al.}{2014}]{pre14}
Presotto V., Girardi M., Nonino M., et al. 2014, A\&A, 565, 126

\bibitem[\protect\citeauthoryear{Rosati et al.}{2014}]{ros14}
Rosati P., Balestra I., Grillo C., et al. 2014, Msngr, 158, 48

\bibitem[\protect\citeauthoryear{Trentham \& Hodgkin}{2002}]{tre02}
Trentham N. \& Hodgkin S., 2002, MNRAS, 335, 712
}
\end{thebibliography}
\end{document}